\documentclass[epj,twocolumn]{svjour}
  
\usepackage{amsmath,amssymb}
\usepackage{graphicx}
\usepackage{cite}
\usepackage{color}
\journalname{Eur. Phys. J. B}

\graphicspath{{plots/}{img/}}

\newcommand{\g}{{\vec g}}
\renewcommand{\k}{{\vec{k}}}

\newcommand{\sgn}{\mathop{\mathrm{sgn}}}

\begin{document}

\title{Graphene under bichromatic driving: Commensurability and spatio-temporal symmetries}
\titlerunning{Graphene under bichromatic driving}
\author{Sigmund Kohler and Tobias Stauber}
\institute{Instituto de Ciencia de Materiales de Madrid, CSIC, E-28049 Madrid, Spain}

\date{\today}


\abstract{We study the non-linear current response of a Dirac model that is
coupled to two time-periodic electro-magnetic fields with different
frequencies.  We distinguish between incommensurable and commensurable
frequencies, the latter characterized by $\Omega_2=(p/q)\Omega_1$ with
co-prime integers $p$ and $q$.  Coupling the (effective) two-level system
to a dissipative bath ensures a well-defined long-time solution for the
reduced density operator and, thus, the current. We then analyze the
spatio-temporal symmetries that force certain current components to vanish
and close with conclusions for directed  average currents.   
}

\maketitle

\section{Introduction}

Driven systems are ubiquitous in solid-state physics\cite{Grifoni1998a,
Forster2015b, Kohler2005a}, and recently their relation to emergent
topological phases has attracted much attention
\cite{Kitagawa10,Lindner11,Gu11,Jiang11,Gomez13,Cayssol13}. Remarkably, a
non-trivial Berry curvature can also induce a temporal phase transition if
the driving is sufficiently strong \cite{Rudner19} and driving protocols
can further be extended to quantum dot arrays that might be used as quantum
simulators for 1D topological phases \cite{PerezGonzalez19}. This opens up
new possibilities for manipulating states of matter via strong external,
time-periodic fields. 
For a two-level system with Chern number $C$, a quantized energy pumping
with rate $(C/2\pi)\omega_1\omega_2$ can then occur if the two frequencies
$\omega_1$ and $\omega_2$ are incommensurable \cite{Martin17}.

An established technique for treating ac driven systems beyond linear
response is Floquet theory which is usually employed for simple harmonic
time-dependences.  Nevertheless, multi-frequency driving with
incommensurable frequencies can be treated as well, but requires a
multi-dimensional Floquet lattice \cite{Hanggi1998a}. Moreover, one can map
such quasi-periodic systems to periodic systems leading to Floquet time
spirals \cite{Zhao19}.

In this paper, we shall investigate a two-level system driven by two
frequencies that may be commensurable as well as incommensurable. The
system is further coupled to a dissipative decay channel which leads to a
master equation description for the reduced dynamics. Its long-time
solution provides the time averaged current. We are then interested in
analyzing the symmetries of the system that lead to a vanishing current.

The creation of directed currents by purely oscillating forces has a long
history in classical Brownian motion, where it is known as ratchet effect
\cite{Reimann2002a, Hanggi2009a}. It has been studied also in the quantum
realm \cite{Reimann1997a, Lehmann2002b}.  Generally, ratchet effects stem
from an interplay of non-linearities and ac driving that brings the system
out of equilibrium such that detailed balance is broken. Symmetries that
inhibit the ratchet effect usually take the shape of the driving as a
function of time into account, i.e., they are of spatio-temporal nature
\cite{Flach2000a, Reimann2001a}.  Lately, such concepts have been developed
also in the context of Floquet topological insulators \cite{Peng2019a}.

The paper is organised as follows: In Sec.~\ref{sec:Model}, we will
introduce the Dirac model and Chern number and relate the two-level system
to various physical systems. In Sec.~\ref{sec:Solution}, we will then
outline the master equation and the Floquet techniques for its solution.
In Sec.~\ref{sec:SpatioTemporalSym}, we discuss the conditions
for a vanishing current in a certain direction and in
Sec.~\ref{sec:numerics}, we present the numerical results resolved in
$k$-space. We close with conclusions and final remarks.

\section{Model}
\label{sec:Model}
\subsection{Graphene in two electromagnetic fields}

The motivation of this work is to discuss the non-linear current response
of graphene, but in order to simplify the general discussion, we will
constrain ourselves to only one valley. The Hamiltonian is thus given by (in
units with $\hbar=v_F=e=1$)
\begin{equation}
H_0({\vec k}) = \vec k\cdot {\vec\sigma} + m\sigma_z,
\quad {\vec\sigma} = \begin{pmatrix}\sigma_x\\\sigma_y \end{pmatrix}\;,
\end{equation}
with the gap parameter $m$ and the quasi momentum $\vec k$ measured
relatively to the $K$-point.

Via minimal coupling, two orthogonal driving fields are introduced as the
time-dependent Hamiltonians
\begin{align}
\label{drive1}
H_1(t) ={}& a_1 \sigma_x \cos(\Omega_1 t) , \\
\label{drive2}
H_2(t) ={}& a_2 \sigma_y \cos(\Omega_2 t + \theta) ,
\end{align}
where the frequency ratio $\Omega_2/\Omega_1$ may be irrational or
rational, i.e., equal to $p/q$ with co-prime integers $p$ and $q$.  Without
loss of generality, we allow for a phase shift in the second field, only.

The total Hamiltonian thus reads $H(t)=H_0+H_1(t)+H_2(t)$, and we assume that for a
chemical potential at the Dirac point, the two single-particle states for a
given momentum $\vec k$ are fully occupied and empty, respectively, such
that the current density reads
\begin{equation}
\vec j=\frac{1}{A}\sum_\k \langle{\vec\sigma}\rangle
= \frac{1}{4\pi^2} \int d^2k\, \langle\vec\sigma\rangle \;,
\label{current}
\end{equation}
where $\langle\vec\sigma\rangle$ denotes the time-averaged expectation value.

For a realistic description of many-body effects, this one-particle
approach may represent a severe limit.  However, within the present work we
restrict ourselves to analyzing the spatio-temporal symmetries of
single-particle states in an idealized situation.

\subsection{General two-level systems and Chern number}
Our approach can be applied to any two-level system and the general
Hamiltonian defined on a two-dimensional torus would read
\begin{align}
H=\g(\k)\cdot{\vec \sigma}\;.
\end{align}
For $\g(\k)=(\tau k_x,k_y,m)$, this reduces to the Dirac Hamiltonian, i.e.,
single-valley gapped graphene with $\tau=\pm$ denoting the different
valleys \cite{Neto09,Xiao10}. As already said, this Hamiltonian will  be
treated in detail below.

For $\g(\k)=(k_x^2-k_y^2,2\tau k_xk_y,\Delta)$, we would model biased
bilayer graphene \cite{Li10}, resembling one of the first examples of
topological edge states localized in the region for which a sign change in the bias voltage occurs\cite{Martin08}. Setting $\g(\k)=(v_x\sin(k_x), v_y\sin(k_y),
m-b_1\cos(k_x)-b_2\cos(k_y))$, one obtains a half of the BHZ model
\cite{Bernevig06}. The last version was discussed in Ref. \cite{Martin17}
after coupling it to two independent driven fields.

The Chern number of a two-level system in two dimension can be defined as
\begin{align}
C=\frac{1}{4\pi}\int d^2k \ \hat\g\cdot(\partial_{k_x}\hat\g\times\partial_{k_y}\hat\g)\;,
\end{align}
with $\hat\g=\g/|\g|$. For gapped graphene, $C=\tau \sgn(m)$, while for
biased bilayer $C=\tau \sgn(\Delta)$ and for the half BHZ model we have a
quantum Hall insulator with $C=\pm1$ for $-|b_1|-|b_2|<m<-||b_1|-|b_2||$
and $||b_1|-|b_2||<m<|b_1|+|b_2|$, respectively. 

Coupling the graphene Hamiltonian to a circularly polarized light field
\begin{align}
H_\mathrm{circ}(t) = g(\sigma_+e^{i\Omega_1t}+\sigma_-e^{-i\Omega_1t})
\end{align}
may lead to non-trivial topological properties, i.e., for the Dirac
Hamiltonian, we obtain $C=\sgn(m+g)$ for $\tau=+$ and $C=-\sgn(m-g)$ for
$\tau=-$ \cite{Cayssol13}. The net Chern number can thus become non-trivial even after
including both valleys if the coupling is sufficiently strong or the gap
sufficiently small, i.e., $|g|\geq|m|$. 

In addition to the Chern, we could also calculate the more general
dynamical conductivity tensor defined as 
\begin{align}
\sigma_{nm}(\omega)=\frac{1}{\omega}\int_0^\infty e^{i\omega t}\langle[j_n(t),j_m(0)]\rangle\;,
\end{align}
with $n,m=x,y$. In the static limit $\omega\rightarrow0$, we then have the relation
\begin{align}
\sigma_{xy}=\frac{e^2}{2\pi\hbar}C\;,
\end{align}
where we have restored SI-units for the moment. The above equation
resembles the main result of the celebrated integer Hall effect
\cite{Klitzing80,Thouless82,Streda82}. 	

In the following, we will go beyond linear response theory and discuss the
current in the non-linear regime. As a special case, we also analyze the
dynamical Hall response.

\section{Long-time solution}
\label{sec:Solution}
\subsection{Quantum dissipation}

To obtain a well-defined steady state, we introduce a weak dissipation
mechanism.  To this end, one may start from a system-bath model to obtain
an equation of motion for the reduced density operator of the dissipative
system.  Then one can show that generally dissipation is quantitatively
affected by the driving \cite{Kohler1997a, Grifoni1998a}.  Here, however,
we are interested in the generic response to bichromatic driving and we
will follow a less involved path which allows an efficient numerical
solution for rather long propagation times.  Therefore, we simply employ a
Lindblad master equation for the density operator \cite{Breuer2003a},
$\dot\rho = -i[H_0+H_1+H_2, \rho] + \gamma\mathcal{D}(\rho)$ with the
Lindblad dissipator \cite{Breuer2003a}
\begin{equation}
\mathcal{D}(\tilde\sigma_-)\rho
= \tilde\sigma_-\rho \tilde\sigma_+^\dagger
  -\frac{1}{2}\tilde\sigma_+^\dagger \tilde\sigma_-\rho
  -\frac{1}{2}\rho \tilde\sigma_+^\dagger \tilde\sigma_- ,
\end{equation}
where $\tilde\sigma_-= |\varphi_0\rangle\langle\varphi_1|$ is the ladder
operator in the eigenbasis of $H_0$ which maps the excited state to the
ground state.

\subsection{Solution of the master equation}

For time-dependent master equations of this type, the time-averaged
long-time solution of the density operator can be obtained with Floquet
methods, both in the commensurable and the incommensurable case.  In the
following, we sketch the underlying ideas and for details refer the reader
to Ref.~\cite{Forster2015b}.

\subsubsection{Commensurable frequencies}
\label{sec:comm}

For $\Omega_2 = (p/q)\Omega_1$, with co-prime $p$ and $q$, the system is
periodic with a fundamental frequency $\Omega = \Omega_1/q$. Then, since the
master equation is linear and (generally) ergodic, $\rho(t)$ becomes
$2\pi/\Omega$ periodic  after a transient stage and can be written as a
Fourier series
\begin{equation}
\rho(t) = \sum_k e^{-ik\Omega t} \rho_k .
\end{equation}
Inserting this Floquet ansatz into the master equation and choosing a
suitable cutoff for the Fourier index $k$, yields a set of linear equations
for the coefficients $\rho_k$ which we solve numerically.  We are finally
interested in the time average over one period given by $\rho_0$.

\subsubsection{Incommensurable frequencies}
\label{sec:incomm}

For incommensurable frequencies, i.e., for irrational values of
$\Omega_2/\Omega_1$, one may decompose the long-time solution of $\rho(t)$
into a two-dimensional Fourier series, one for each frequency
\cite{Hanggi1998a, Chu2004a}.  This however may lead to rather large sets
of equations which are hard to solve numerically.  For a more efficient
treatment, we employ a method \cite{Forster2015b} based on the combination
of the Floquet decomposition explained above, the $t$-$t'$ formalism
\cite{Peskin1993a}, and matrix-continued fractions \cite{Risken}.

The method starts by replacing one time argument in the Liouvillian
by $t'$ to obtain the modified master equation
\begin{equation}
\Big(\frac{\partial}{\partial t} + \frac{\partial}{\partial t'}\Big)
R(t,t') = \mathcal{L}(t,t') R(t,t')
\end{equation}
with
\begin{equation}
\mathcal{L}(t,t') = L_0 + L_1\cos(\Omega_1 t) + L_2\cos(\Omega_2 t' + \theta) .
\label{tt'}
\end{equation}
It can be shown straightforwardly that when $R(t,t')$ is a solution of
Eq.~\eqref{tt'}, then $R(t,t')|_{t'=t}$ solves the original master equation
\cite{Peskin1993a}.  Practically, one treats the additional time $t'$ (or
equivalently, some angle $\alpha = \Omega_2 t'$) as additional canonical
coordinate with conjugate momentum $i\partial/\partial t$.

For Eq.~\eqref{tt'}, the ansatz
\begin{equation}
R(t,t') = \sum_{k,n} e^{-ik\Omega_1 t} e^{-in\Omega_2 t'} r_{k,n} \,,
\end{equation}
yields a set of equations which is tri-diagonal in both indices, $k$ and
$n$.  It is solved by writing the dependence on one index as a Floquet
matrix like in Sect.~\ref{sec:comm} while the dependence on the other index
is expressed as a recurrence relation that can be solved by
matrix-continued fractions.  The latter numerical method scales only
linearly with the cutoff index, which makes the method considerably more
efficient than the direct matrix representation of the two-frequency
decomposition.  Finally, we obtain the coefficient $r_{0,0}$ which contains
the full information about the long-time average of $\rho(t)$.

It has been shown \cite{Forster2015b} that $r_{0,0}$ does not depend on the
relative phase of the drivings, $\theta$.  This is indeed expected from
physical intuition, because for quasi-periodic driving fields, any relative
phase between the two ac signals can be mapped to a time translation which
should not affect long-time averages.  Below we derive this phase
independence more formally within a symmetry analysis.

\section{Spatio-temporal symmetries}
\label{sec:SpatioTemporalSym}

The first term of the time-independent Hamiltonian $H_0$, i.e., $\vec
k\cdot\vec\sigma$, is given by a inner product which is invariant under
time-reversal (which changes the sign of both $\vec k$ and $\vec\sigma$)
and under a rotation around the $z$-axis.  However, as $m$ and $\sigma_z$
are not components of a vector, a rotation around any other axis does not
correspond to a transformation in real space.  Nevertheless, such rotations
may be symmetry operations for $H(t)$ and must be considered.  For this
reason, we treat in our symmetry analysis $\vec k$ as well as the driving
amplitudes $a_{1,2}$ as parameters that are not affected by the
transformations.  Notice that this does not imply any restriction, because
a possible sign in $k_x$ or $k_y$ is irrelevant for the integral in
Eq.~\eqref{current}, while possible minus signs of the amplitudes can be
absorbed by the relative phase of the driving fields.

The principal observable for our setup is the current density $\vec j$
which is given by an integral of the time-averaged expectation value
$\langle\vec\sigma\rangle$, see Eq.~\eqref{current}.  Thus, whenever a
component of this quantity possesses some anti-symmetry as a function of
$\vec k$, the corresponding current component will vanish.  The aim of this
section is a symmetry analysis of $H(\vec k,t)$ in the spirit of
Refs.~\cite{Flach2000a, Reimann2001a} that reveals under which conditions
one or both current components are symmetry forbidden.  In doing so, we
consider spatio-temporal transformations that map $H(\vec k,t)$ to some
$H(\vec k')$, where $\vec k$ and $\vec k'$ are related by a mirror or point
symmetry.  The spatial part of the mapping is formally a rotation or
inversion in three-dimensions with the corresponding transformation of the
Pauli matrices.

\subsection{Commensurable frequencies}
\label{sec:symm_comm}

\subsubsection{Periodicity in the phase $\theta$}

Before considering transformations of Pauli matrices, let us derive for
later use a symmetry property for the phase $\theta$ of the driving defined
in Eq.~\eqref{drive2}.  Obviously, $H_2(t)$ is $2\pi$ periodic in $\theta$.
In the long-time limit, however, time-averaged expectation values as a
function of $\theta$ possess a higher symmetry, namely a $2\pi/q$
periodicity which we derive in the following.

As already mentioned above, for rational $\Omega_2/\Omega_1 = p/q$, the
Hamiltonian is periodic in time.  Then after a transient stage, the density
operator $\rho(t)$ generally assumes the same time periodicity, and so does
any expectation value of a time-independent observable.\footnote{Exceptions
are typically found for somewhat artificial models in which both the bath
coupling and the driving commute with $H_0$.} Hence, all averages over
one driving period are invariant under time translations.  We are now
interested in phase transformations $\theta \to \theta + \Delta\theta$ with
a $\Delta\theta$ that can be absorbed into a time translation $\tau$, such
that averages over one driving period remain invariant.

From the definition of $H_2(t)$, we immediately see that such a phase shift
corresponds to a time translation by $\tau = \Delta\theta/\Omega_2$.  This,
in turn, provides for the driving $H_1$ a phase shift
$\Delta\theta\Omega_1/\Omega_2 = \Delta\theta q/p$.  Whenever this phase is
a multiple of $2\pi$, $\Delta\theta$ will not affect stationary expectation
values.  This is the case for $\Delta\theta = 2\pi p\ell/q$, where $\ell$
may be any integer.

We choose $\ell = p^{\varphi(q)-1}$, with $\varphi$ being Euler's totient
function which counts the natural numbers up to $q$ that are co-prime to
$q$.  As 1 is considered co-prime to all natural numbers, $\varphi(q)\geq 1$
which ensures that the chosen $\ell$ is an integer number.  As $p$ and $q$
are co-prime, Euler's theorem states that $p^{\varphi(q)} \equiv 1 \mod q$.
Hence, for the present choice,
\begin{equation}
\Delta\theta = \frac{2\pi}{q} ,
\label{Deltatheta}
\end{equation}
which implies the to be demonstrated $2\pi/q$ periodicity of time-averaged
expectation values.

\subsubsection{Temporal symmetries of the driving shape}

Next we consider the time dependent functions in the driving Hamiltonians
$H_{1,2}(t)$ given by
\begin{align}
\label{f1}
f_1(t) ={}& \cos(q\Omega t) , \\
\label{f2}
f_2(t) ={}& \cos(p\Omega t + \theta) .
\end{align}
We are interested in transformations that change the sign of at least one
of these functions and accordingly classify them by $\vec s = (s_1,s_2)$
with $s_i=\pm$.  For the cosine, two transformations come to mind.  First,
a time translation and, second, time-reversal at times that correspond to
zeros of $f_{1,2}$.  Importantly, due to the periodicity worked out above,
we have the freedom to change $\theta$ by any multiple of $2\pi/q$ without
affecting the time-averaged response.

\paragraph{Time translation}
The first option is the mapping $t\to t+\tau$, where $\tau$ will be
determined such that $f_1$ acquires a sign $s_1$.  Thus, $\tau =
2\pi\ell/q\Omega$ for $s_1=+$ and $\tau = \pi(2\ell+1)/q\Omega$ for
$s_1=-$, where $\ell$ is an arbitrary integer.  The corresponding condition
on $f_2$ reads
\begin{equation}
\cos(p\Omega t+\theta) = s_2 \cos(p\Omega t+p\Omega\tau+\theta')
\end{equation}
and must be fulfilled for all $t$, while $\theta' \equiv
\theta\pmod{2\pi/q}$ owing to the aforementioned $2\pi/q$ periodicity.
Inserting the already determined values of $\tau$ straightforwardly leads
to the conditions summarized in the first line of
Table~\ref{tab:cossymmetry}.  Notice that a minus sign in $f_2$, i.e.\ a
phase $\pi$, can be absorbed by $\theta'$ provided that $q$ is even.

\paragraph{Time reversal}
As $t$ enters as argument of the cosines, its sign is irrelevant for $f_1$.
For $f_2$, the mapping $t\to -t$ is equivalent to changing the sign of
$\theta$.  Allowing again also an additional time translation and a phase
shift by a multiple of $2\pi/q$, the most general time inversion reads
$(t,\theta) \to (t+\tau,-\theta')$.  Then, $f_1$ is not affected such that
we find for $\tau$ the same possible values as above.  The difference lies
in the minus sign in front of $\theta'$ such that for $\vec s=(-+)$, the
condition on the arguments of $f_2$ becomes
\begin{equation}
\theta \equiv \frac{\pi p}{q}(2\ell + 1) -\theta \mod \frac{2\pi}{q} ,
\end{equation}
where we have used $\theta \equiv \theta' \pmod{2\pi/q}$.
Thus, $\theta$ no longer disappears from the symmetry condition, but for
even $q$ must be $\theta\equiv 0 \pmod{\pi/q}$.  For odd $q$, it is
restricted to $\theta\equiv \pi/2q \pmod{\pi/q}$.  Notice the absence of
the factor $2$ in the modulus.  The conditions for $\vec s=(-+)$ and $\vec
s=(--)$ are evaluated in the same manner and provide the second and third
line of Table~\ref{tab:cossymmetry}.

\begin{table}[t]
\centering
\caption{Symmetries of the driving under time translation $t\to t+\tau$ and
time reversal $t\to -(t+\tau)$.  The signs of the driving shapes $f_1(t) =
\cos(q\Omega t)$ and $f_2(t) = \cos(p\Omega t+\theta)$ change as indicated
in the first line provided that $p$, $q$, and $\theta$ obey the conditions
listed in the subsequent lines.  Notice that as $p$ and $q$ are co-prime by
assumption, $p+q$ is even only when both $p$ and $q$ are odd.
\label{tab:cossymmetry}}
\begin{tabular*}{\columnwidth}{@{\extracolsep{\fill}}lccc}
\hline
Sign change $\vec s$ of $f_1$, $f_2$ & $+-$ & $-+$ & $--$ \\
\hline
Time translation & $q$ even & $p$ even & $p+q$ even \\
Time reversal, $\theta=0$ & $q$ even & $p$ even & $p+q$ even \\
Time reversal, $\theta=\pi/2q$ & $q$ odd & $p$ odd & $p+q$ odd \\
\hline
\end{tabular*}
\end{table}

\subsubsection{Transformation of the Pauli matrices}

The spatial part consists of the usual behavior of Pauli matrices under
rotation and time reversal \cite{Sakurai}. Due to the fact that a symmetry
transformation must not mix the couplings to the ac drivings (unless
$\Omega_1=\Omega_2$, the only possibilities are transformations that change
the sign of one or several Pauli matrices.  They are given by combinations
of rotations at the coordinate axis by an angle $\pi$ and time reversal,
where all $2^3-1$ possibilities (the identity is not relevant for our
purpose) are listed in the top row of Table~\ref{tab:symmetry}.  

Let us once more emphasize that we consider the momentum $\vec k$ and the
amplitudes $a_1$, $a_2$ as mere parameters, such that $T$ only acts on the
(pseudo)-spin space and the arguments of the cosines.  Their action on the
Pauli matrices is displayed in the second row of Table~\ref{tab:symmetry}.

\begin{table*}[t]
\centering
\caption{Spatio-temporal symmetries for which the current $\vec j =
(j_x,j_y)$ or one of its components are zero.  They include all possible
combinations the time-reversal $T$ and rotations $R_i$ by an angle $\pi$ at
the coordinate axes $i=x,y,z$.  The conditions on the phase $\theta$ are
modulo $\pi/q$ and, thus, in the range $[0,2\pi)$ are fulfilled $2q$ times.
The last two columns list the most relevant symmetries as for those the
entire current vanishes.  The mapping via $TR_z$ solely affects $\sigma_z$
which is not linked to any current.
\label{tab:symmetry}}
\begin{tabular*}{\textwidth}{@{\extracolsep{\fill}}lccccccc}
\hline
Symmetry operation & $TR_x$ & $TR_y$ & $TR_z$ & $R_x$ & $R_y$ & $R_z$ & $T$ \\
\hline
Signature, impact on $\sigma_x,\sigma_y,\sigma_z$
 & $-++$ & $+-+$ & $++-$ & $+--$ & $-+-$ & $--+$ & $---$ \\
\hline
Restrictions: \\
-- gap & & & $m=0$ & $m=0$ & $m=0$ & & $m=0$ \\
-- frequency ratio $\Omega_2/\Omega_1=p/q$ &&&& $q$ even & $p$ even & $p,q$ odd & \\
-- phase $\theta \pmod{\pi/q}$ &
    $\begin{cases} 0 & \text{$p$ even} \\ \frac{\pi}{2q} & \text{$p$ odd} \end{cases}$ &
    $\begin{cases} 0 & \text{$q$ even} \\ \frac{\pi}{2q} & \text{$q$ odd} \end{cases}$
    &&&&&
    $\begin{cases} 0 & \text{$p+q$ even} \\ \frac{\pi}{2q} & \text{$p+q$ odd} \end{cases}$ 
\\[2.8ex]
\hline
Consequence for current & $j_x=0$ & $j_y=0$ & & $j_y=0$ & $j_x=0$ & $\vec j=\vec 0$ & $\vec j=\vec 0$
\\
\hline
\end{tabular*}
\end{table*}

\subsubsection{Combining both transformations}

Armed with the knowledge of the previous subsections, we are in the
position to analyze the spatio-temporal symmetries of our problem.
Notably, the spatial transformations in Table~\ref{tab:symmetry} invert the
sign of at least one Pauli matrix.  Then, under the conditions listed in
Table~\ref{tab:cossymmetry}, there exists a time transformation that
restores the original sign of the driving Hamiltonians $H_1$ and $H_2$.
Thus, the combination of both transformations maps $H(\vec k)$ to some
$H(\vec k')$ with $\vec k' = (\pm k_x,\pm k_y)$.  Therefore owing to the
integration in Eq.~\eqref{current}, a current component $j_i$ vanishes if
the transformation inverts the sign of $\sigma_i$.

An important point is that $\sigma_z$ does not couple to any driving field
nor does it depend on the momentum.  Therefore, any mapping that involves
$\sigma_z\to-\sigma_z$ can be a symmetry operation only in the gapless case
$m=0$.

As an example, let us consider a $\pi$-rotation around the $z$ axis, which maps
$H_0(\vec k)$ to $H_0(-\vec k)$ and $H_{1,2}(t) \to -H_{1,2}(t)$.
According to the last column of Table~\ref{tab:cossymmetry}, there exists
for even $p+q$ (i.e., $p,q$ both odd) a time translation that restores the
sign of both drivings.  Therefore, we can conclude that the momenta $\vec
k$ and $-\vec k$ are symmetry related.  As both $\sigma_x$ and $\sigma_y$
have changed their sign, the contributions of these momenta to the current
$\vec j$ cancel each other.  Thus, the current is symmetry forbidden.  Let
us remark that this case represents the most important symmetry: first,
because both current components vanish and, second, as it holds also in the
gapped case.

Generally, for any spatial transformation, one has to look at
Table~\ref{tab:cossymmetry} for a time transformation with the same
signature (ignoring the last one which corresponds to $\sigma_z$), such
that the driving Hamiltonians remains invariant.  Then all $\sigma_i$ that
change their sign possess some anti-symmetry in $\vec k$ space.  Hence,
$j_i=0$.  The conditions under which a proper time transformation exists
can be read off from Table~\ref{tab:cossymmetry} and provide the
restrictions on $p$, $q$, and $\theta$ displayed in
Table~\ref{tab:symmetry}.

Notice that in some cases, a symmetry may represent a special case of a
higher symmetry.  For example, time reversal symmetry $T$ predicts for the
gapless case $m=0$ and $p+q \equiv 0 \mod{2}$ (implying that both $p$ with
$q$ are odd) a vanishing current for particular values of $\theta$.  For this
case, however, the rotation around the $z$-axis is less restrictive and
leads to the same conclusion for any phase and even for a finite gap.

\subsubsection{Generalization of the Hamiltonian}
\label{sec:generalization}

Rotation around the $z$ axis, $R_z$, as well as time reversal $T$ have in
common that both Pauli matrices relevant for the current, $\sigma_x$ and
$\sigma_y$, transform in the same way.  Therefore, in the driving
Hamiltonians we may replace $\sigma_x$ and $\sigma_y$ by any linear
combination of the two without loosing the corresponding symmetry
properties.  Physically, this means that the polarization of the two
incident electric fields need not be orthogonal, but may have any
orientation in the $x$-$y$ plane.  Interestingly, this property holds true
precisely for those symmetries for which both current components vanish.

For the other symmetries (besides for $TR_Z$ which does not have
consequences for the current), this generalization is not possible, because
the spatial part of the transformation changes the sign of only one of the
two Pauli matrices that define the current.

\subsection{Incommensurable frequencies}

For irrational $\Omega_2/\Omega_1$, the system always possesses the
highest symmetry that can be achieved in the commensurable case.  This can
be understood as follows.  The symmetry analysis for the commensurable case
is based on the compensation of a prefactor $-1$ in the driving by a
proper time transformation.  For an arbitrary frequency $\Omega_1$, the sign
of $H_1$ can be inverted by a time translation $t\to t
+\pi(2\ell+1)/\Omega_1$ with an arbitrary integer $\ell$.  Then, the phase
in $H_2$ effectively changes by
\begin{equation}
\Delta\theta = \pi \frac{\Omega_2}{\Omega_1}(2\ell + 1) \mod 2\pi .
\end{equation}
If $\Omega_2/\Omega_1$ is irrational, one can always choose $\ell$ such
that it brings $\Delta\theta$ arbitrarily close to its original value (or
to any other desired values, e.g., to $\pi$ if one wishes to establish a
minus sign).  By contrast, for $\Omega_2/\Omega_1 = p/q$ with $p,q$ being
co-prime, the possible phase shifts can assume only $q$ different values.

Notice that this argument silently assumes that the average is computed
over an infinitely large time.  Therefore, it will be difficult to
distinguish in an experiment an incommensurable case from a commensurable
case with rather large $p$ and $q$.  

The arguments used in Sec.~\ref{sec:generalization} for the generalization
of the driving also hold here.  Therefore, also for incommensurable
frequencies, we can replace in $H_1$ and $H_2$ the Pauli matrices by any
linear combination of $\sigma_x$ and $\sigma_y$ without loosing the
symmetry properties that lead to a point symmetry of
$\langle\sigma_{x,y}\rangle$ with respect to the origin of $k_x$-$k_y$ plane and, thus,
to a vanishing current density.

\begin{figure*}
\centerline{\includegraphics{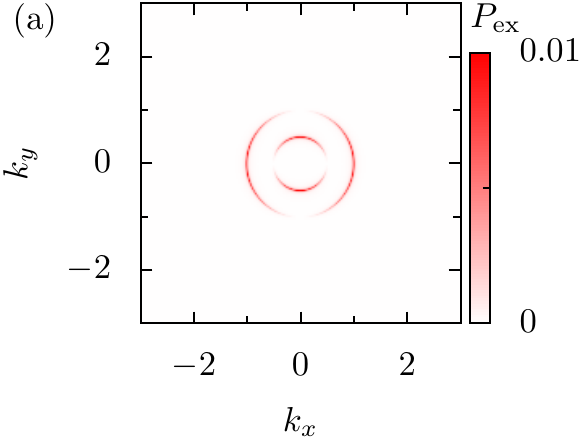}
\includegraphics{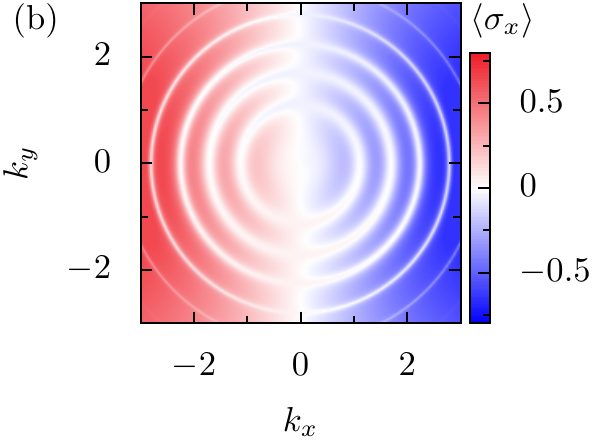}}
\caption{(a) Excitation probability as function of the momentum $\vec k$
for driving frequencies $\Omega_2=2\Omega_1$ and equal amplitudes $a_1=a_2
= 0.01$ within the linear response regime.  
(b) Time-averaged $\langle\sigma_x\rangle$ for monochromatic driving
$\Omega_2=\Omega_1$ with a phase $\theta=\pi/2$ that corresponds to
circular polarization.  The much larger amplitudes $a_1=a_2=3$ create
a response dominated by higher harmonics. For illustrative purposes, the
dissipation rate is chosen rather large, $\gamma=0.05$.
}
\label{fig:monochrome}
\end{figure*}
\begin{figure*}
\centerline{ \includegraphics{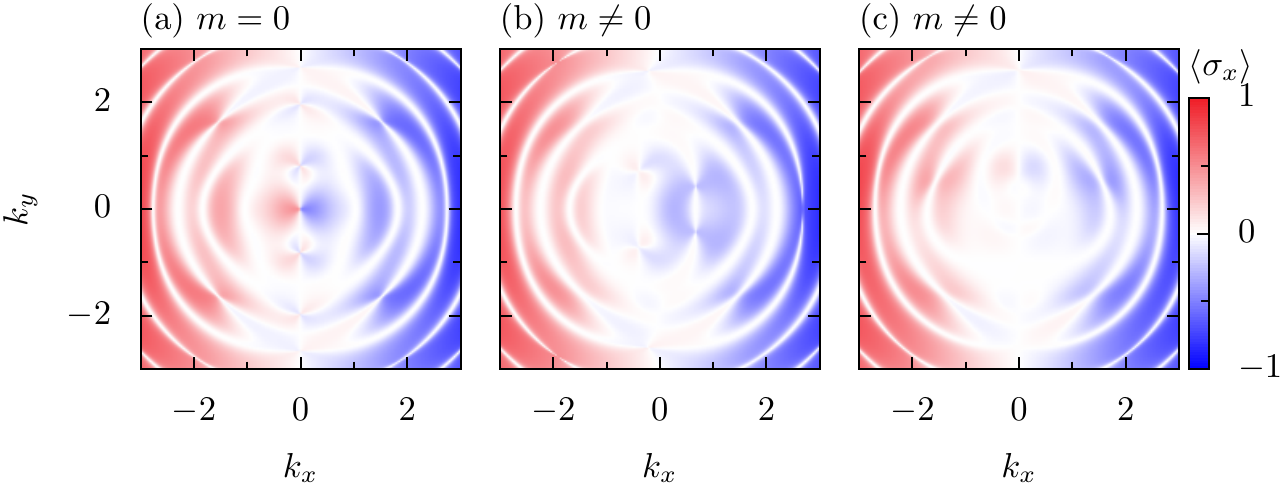} }
\caption{Expectation value $\langle\sigma_x\rangle$ as a function of
momentum $\vec k$ for the resonance $p/q=2/3$ with $\Omega_1=1$.
(a) In the absence of a gap ($m=0$) for the phase $\theta=\pi/2q$.
(b) For a gap $m=0.5$ and phase $\theta=\pi/2q$.
(c) For a gap $m=0.5$ and phase $\theta=0$.
The amplitudes are $a_1=a_2=3$, while the dissipation rate is
$\gamma=10^{-3}$.}
\label{fig:pq23}
\end{figure*}

\subsection{Hall response}

By setting one driving field to zero, we can also discuss the possibility
of a non-linear Hall response of the system, i.e., we are looking for a current in,
say, $y$-direction if the external driving field is applied in $x$-direction. 

One might expect some non-trivial response in the case of a finite gap, 
but since there is only one driving field, the highest possible symmetry is attained
by the system. The highest symmetry  class is also represented by both $p$ and $q$
odd and we infer from Table~\ref{tab:symmetry} that the total current is
always zero and thus no non-linear Hall current can be generated. This is true for any
finite frequency and there is thus no dynamical Hall effect induced by non-linear radiation. 
\section{Numerical results}
\label{sec:numerics}

Besides the current densities $j_{x,y}$, our main quantity of interest are
the time-averaged expectation values of the Pauli matrices $\sigma_{x,y}$
as a function of $\vec k$.  Both are linked by the integral in
Eq.~\eqref{current}.  Physical insight may also be provided by the
probability of finding the system in the excited state.  To compute the
latter, we determine the excited state $|\phi_1\rangle$ of the undriven
$H_0(\vec k)$ (the index $\vec k$ in the energies and eigenstates is
suppressed).  Then we evaluate $P_\text{ex} =
\langle\phi_1|\rho_\infty|\phi_1\rangle$, where $\rho_\infty$ in the
one-period average of the density operator or, in the incommensurable case,
its long-time average.

\subsection{Commensurable frequencies}

To set the stage, we first consider the excitation probability for
bichromatic driving with small amplitudes, such that the parameters stay
within the linear response limit, see Figure~\ref{fig:monochrome}a.  As is
characteristic for linear response, there emerge two independent
excitations, one for each driving frequency.  Their shape as a ring
reflects the rotational symmetry of $H_0(\vec k)$.  The zeros of the
excitations at $k_x=0$ and $k_y=0$, respectively, are due to the fact that
for these momenta, one of the drivings commutes with the bare Hamiltonian
$H_0$.  Hence it cannot cause any excitation. The displayed color-coded
intensities are $\propto\cos^2\theta$.

Another instructive case is monochromatic driving with circular
polarization shown in Figure~\ref{fig:monochrome}b, but now with a much
larger amplitude far beyond linear response.  Owing to the equal amplitudes
and the circular polarization, the driving still possesses the rotational
symmetry of $H_0$.  An interesting feature is the counter-clockwise
smearing, which is a consequence of dissipation.  For the opposite circular
polarization, the smearing is clockwise (not shown).  To make this effect
visible, we here used an unphysically large dissipation. In all other
figures it is much smaller such that dissipative effects are not
significant.

To see how symmetries may be destroyed by the presence of a gap and be
restored by choosing a proper phase between the two drivings, let us have a
closer look at a resonance with $p/q=2/3$, for
zero gap, $m=0$, and phase $\theta=\pi/2q$, see Figure~\ref{fig:pq23}.  As $p$ is even, according to
Table~\ref{tab:symmetry}, the system has a symmetry whose spatial part
consists of a rotation by $\pi$ around the $y$ axis, $R_y$.  Consequently,
the time-averaged $\langle\sigma_x\rangle$ as a function of $\vec k$
possesses an anti-symmetry by reflection at the $y$ axis which is evident
in Figure~\ref{fig:pq23}a.  

For finite gap (panel b), this symmetry is no
longer present.  There is also no other symmetry that would affect
$\sigma_x$.  Nevertheless, there is still one symmetry present, namely
$TR_y$ which implies reflection symmetry of $\sigma_x$ at the $x$ axis.
Notice, however, that this has no consequences for the current component
$j_x$, because only anti-symmetries have the effect that the integral in
Eq.~\eqref{current} vanishes.  Upon changing the phase to $\theta=0$ (panel
c), we find invariance under $TR_x$, which for even $p$ has the same
consequence as $R_y$, which is the mentioned anti-symmetry of
$\langle\sigma_x\rangle$.

\subsection{Incommensurable frequencies}

\begin{figure}
\centerline{ \includegraphics{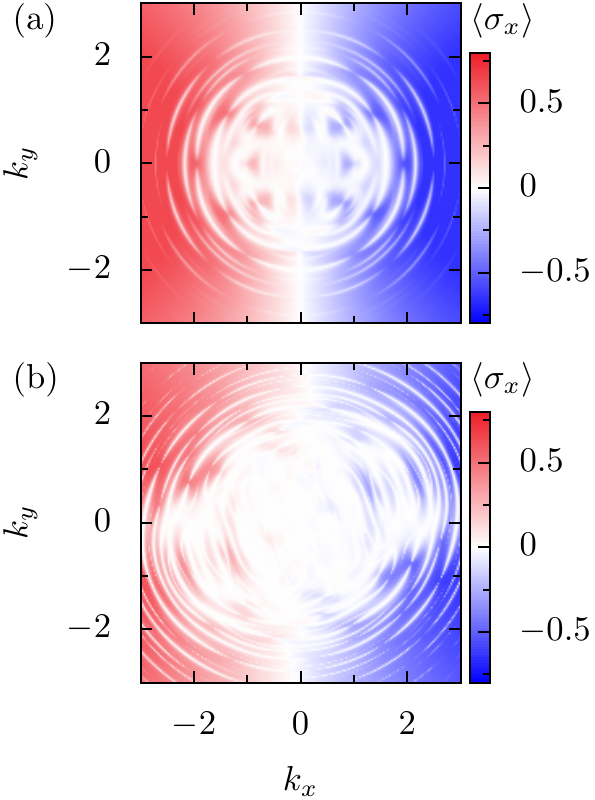} }
\caption{(a) Expectations value $\langle\sigma_{x}\rangle$ for
incommensurable frequencies with $\Omega_2/\Omega_1 = (1+\sqrt{5})/2$,
i.e., equal to the golden ratio.
(b) The same for a modified $H_2$ in which $\sigma_y$ replaced by
$(\sigma_x+\sigma_y)/\sqrt{2}$.
The corresponding plots for $\langle\sigma_y\rangle$, besides a ration by
$90^\circ$, look similar and have the same point symmetry.}
\label{fig:golden}
\end{figure}

As a significant example of incommensurable frequencies, we consider a
frequency ratio equal to the golden mean, which is considered as the ``most
irrational number''.  The resulting expectation value shown in
Figure~\ref{fig:golden}a exhibits many resonance islands without a
particular structure.  On a rough scale, the excitation probability does
not possess any preferential direction despite its lack of rotational
symmetry.  Nevertheless, as both driving fields are orthogonal to each
other, the reflection symmetry at the $k_x$ and $k_y$ axis remains.  For
the expectation values of $\sigma_x$ and $\sigma_y$, this turns into an
anti-symmetry.  Consequently, after integration over $\vec k$ space, the
response vanishes as in the case of commensurable frequencies with a
particular phase.

Figure~\ref{fig:golden}b depicts the corresponding result when the
polarization of the driving $H_2$ is rotated by $45^\circ$, i.e., when
$\sigma_y$ is replaced by $(\sigma_x+\sigma_y)/\sqrt{2}$.  As expected,
then the reflection (anti-) symmetry at the coordinate axis gets lost.
Nevertheless, the point anti-symmetry at the origin still holds, thus again
leading to $\vec j=\vec 0$.

\subsection{Directed average current}

\begin{figure}
\centerline{\includegraphics{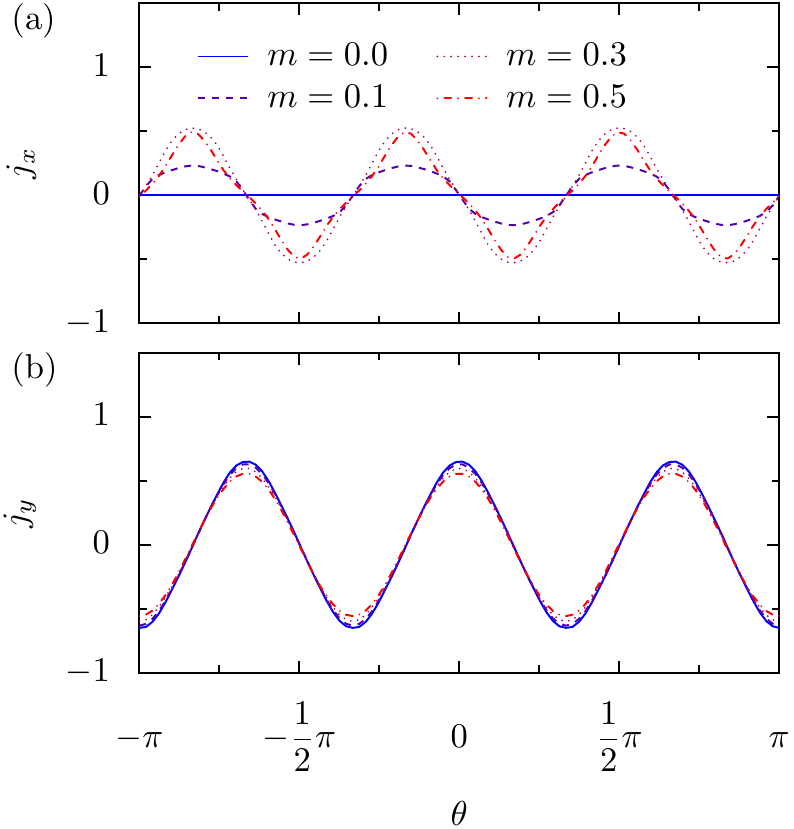}}
\caption{Current for $p/q=2/3$ for various values of the gap.  All other
parameters are as in Figure~\ref{fig:pq23}}
\label{fig:current}
\end{figure}

Let us now discuss an experimentally accessible quantity namely the dc
current.  From our symmetry analysis, we have already seen that it must
vanish when the system is driven by incommensurable frequencies.  The same
is true for commensurable frequencies for $\Omega_2/\Omega_1=p/q$ with both
$p$ and $q$ odd.  Therefore, we focus on cases with either $p$ or $q$ being
even and study the role of the phase $\theta$ between the two driving
fields.

We again consider the case $\Omega_2/\Omega_1 = 2/3$ for which the
time-averaged current is depicted in Figure~\ref{fig:current}.  As $p$ is
even, the ungapped case $m=0$ has the symmetry $R_y$, such that we expect
$j_x$ to vanish.  In the presence of a gap, we may still have a situation
with $TR_x$ and $TR_y$, which however are symmetries only for certain
phases.  Since $p$ is even and $q$ odd, the current components $j_x$ and
$j_y$ vanish for $\theta\equiv 0\pmod{\pi/q}$ and for $\theta\equiv
\pi/2q\pmod{\pi/q}$, respectively.  The numerical data confirm the
conjectured appearance of $2q=4$ zeros of each current component.  While
$j_y$ depends only weakly on the gap, the behavior of $j_x$ changes
significantly.  For $m=0$, it vanishes owing to the discussed invariance of
$H(t)$ under $R_y$.  With increasing $m$, $j_x$ grows until it reaches the
order of magnitude of $j_y$.  When $p$ and $q$ are interchanged (not
shown), the behavior of $j_x$ and $j_y$ is interchanged as well.  The
qualitative difference to the former case is that each current component
vanishes 6 times since now $q=3$.

We have already argued and seen in Figure~\ref{fig:golden} that for
incommensurable frequencies, the symmetry is always the highest one that we
can get in the commensurable case.  Therefore, the current will always
vanish also beyond linear response and at any order.  This summarizes the
predominant consequence of incommensurability in our strongly
bichromatically driven system.

\section{Conclusions}

We have analyzed the Dirac model coupled to the radiation of two
time-periodic fields with different, possibly incommensurable frequencies.
Based on an extensive symmetry analysis, we found that inducing a steady,
long-time current requires a frequency ratio $p/q$ with odd $p+q$ (if
co-prime). Theoretically, the $q$ different equivalent values of $\theta$
may lie so close to each other that they cannot be resolved experimentally,
especially for large $q$.  This limits the possibilities for distinguishing
in an experiment between commensurable and incommensurable frequencies to
clear cases such as the golden ratio or ratios with rather small $p$ and
$q$.

Some points  have been left open. So far, many-body effects due to the
anti-symmetrization of the fermionic wave function have been neglected.
Also, in order to address topological quantities more thoroughly, the
static limit would have to be performed which can be done within the
presented scheme by treating one driving field as perturbation via linear
response. These issues raise intriguing questions for further
investigations.

\begin{acknowledgement}
This work was supported by the Spanish Ministry of Science, Innovation, and
Universities through grants No.\ MAT2017-86717-P and FIS2017-82260-P, as
well as by the CSIC Research Platform on Quantum Technologies PTI-001. It
was initiated at Aspen Center for Physics, which is supported by National
Science Foundation grant PHY-1607611.
\end{acknowledgement}

\bibliographystyle{epj}

\end{document}